\documentclass[12pt]{article}
\usepackage[english]{babel}
\usepackage{amssymb,slashed,latexsym,amsmath,multirow,color}
\usepackage{slashed}
\usepackage{tikz}
\usetikzlibrary{shapes,shadows,arrows}
\pdfoutput=1
\usepackage[font=footnotesize,labelsep=newline,labelfont=sc,justification=centering,position=top]{caption}

\makeatletter
\newcommand*\rel@kern[1]{\kern#1\dimexpr\macc@kerna}
\newcommand*\widebar[1]{%
  \begingroup
  \def\mathaccent##1##2{%
    \rel@kern{0.8}%
    \overline{\rel@kern{-0.8}\macc@nucleus\rel@kern{0.2}}%
    \rel@kern{-0.2}%
  }%
  \macc@depth\@ne
  \let\math@bgroup\@empty \let\math@egroup\macc@set@skewchar
  \mathsurround\z@ \frozen@everymath{\mathgroup\macc@group\relax}%
  \macc@set@skewchar\relax
  \let\mathaccentV\macc@nested@a
  \macc@nested@a\relax111{#1}%
  \endgroup
}
\makeatother

\numberwithin{equation}{section}

\usepackage{hyperref}

\newsavebox{\uuunit}
\sbox{\uuunit}
    {\setlength{\unitlength}{0.825em}
     \begin{picture}(0.6,0.7)
        \thinlines
        \put(0,0){\line(1,0){0.5}}
        \put(0.15,0){\line(0,1){0.7}}
        \put(0.35,0){\line(0,1){0.8}}
       \multiput(0.3,0.8)(-0.04,-0.02){12}{\rule{0.5pt}{0.5pt}}
     \end {picture}}

\def\be{\begin{equation}}
\def\ee{\end{equation}}
\def\ba{\begin{array}}
\def\ea{\end{array}}
\def\bea{\begin{eqnarray}}
\def\eea{\end{eqnarray}}
\def\bd{\begin{displaymath}}
\def\ed{\end{displaymath}}

\begin{document}
\begin{flushright}
\hfill{ \
\ \ \ \ MIFPA-13-05\ \ \ \ }
\end{flushright}
\vskip 1.2cm
\begin{center}
\Large{Finite-temperature R-squared quantum gravity}
\end{center}
\begin{center}
C. D. Burton
\end{center}
\begin{center}
\textsl{George and Cynthia Woods Mitchell Institute for Fundamental Physics and\\
Astronomy, Texas A$\&$M University, College Station, TX 77843, USA}
\end{center}
\begin{center}
\textsl{chris.burton@tamu.edu}
\end{center}
\
\
\begin{center}
ABSTRACT
\end{center}
\begin{sloppypar}
The quantum gravity path integral's measure can be written as the product of classical backgrounds and quantum fluctuations about each background.  After proving that fluctuations about the background do not diffuse in Hilbert space and obey the laws of many-body statistics, their probability distributions, entropy, and expected background are determined.  This background obeys expectation-valued Einstein equations and features an entropy-based positive cosmological constant.  From the fluctuation probability distributions, a finite temperature, R-squared, quantum gravity path integral is constructed whose action presents an interaction picture of quantum gravity that `moves with' the expected background in Hilbert space.  Within this interaction picture of quantum fluctuations about an expected background, the fields required to describe quantum gravity have been transformed into `ordinary' quantum fields propagating on this `rigid' or `fixed' expected background.  Back-reaction has been fully accounted for, and the quantum formulation is manifestly background independent.
\end{sloppypar} 
\section{Introduction}\label{intro}
\begin{sloppypar}
Quantum gravity is a `self-contained' quantum system meaning it does not exist as an ordinary quantum system in a rigid, fixed background.  Metric fluctuations `push' against the background, and the background `pushes back' against the fluctuations.  Quantum gravity is therefore a two-part problem:  Identify the correct background in which all back-reaction has been fully incorporated, and then identify the correct quantum description of fluctuations about this background.  Quantum gravity can now be described with `ordinary' quantum fields propagating on this `rigid' or `fixed' background.        
\end{sloppypar}
\begin{sloppypar}
To identify the correct background, this paper studies (from the viewpoint of many-body statistical physics) the ground state of the quantum vacuum in the absence of matter sources.  All results are derived from one crucial observation:  The zero-point fluctuations of the quantum Riemann tensor are in many-body equilibrium with each other, therefore their probability distributions obey the laws of statistical physics.  The explicit form of these distributions is found by examining the grand-canonical fluctuations of a quantum field which creates and destroys quanta of vacuum, and the resulting fluctuation entropy is a functional of expected curvature.  Because of this functional dependence on expected curvature, entropy maximization requires an Euler-Lagrange variation of entropy with respect to the expected background which produces expectation-valued Einstein equations\footnote{The $(+,-,-,\dotsb)$ and $R_{ab}=R^{c}_{\;\,acb}$ conventions are being used.}:
\begin{equation}\label{MFE}
\langle G \rangle_{\mu\nu} -\Lambda \langle g \rangle_{\mu\nu} =0
\end{equation}
Since dynamical fluctuations have pressure, it is not surprising that the quantum Riemann tensor's zero-point fluctuations would also have pressure:        
\begin{equation}\label{lamda}
\Lambda=\frac{1}{L_G^2}\ln\left(\frac{\sqrt{\pi e}\, \ell_{R}^2}{\ell_0^2}\right)
\end{equation}
This entropy-based $\Lambda$ is explicitly derived along with its various parameters in sections \ref{manybody} and \ref{eom}, and it belongs to the back-reaction category of cosmological constant sources \cite{weinberg, dereview1, DEreview}.  Thermodynamics as a source of dark energy has been discussed in \cite{thermdarkenergy}, and the microscopic formulation of the expectation-valued Einstein equations can be viewed as the microscopic origin of Jacobson's equation of state \cite{jacobson}.                       
\end{sloppypar}
\begin{sloppypar}
In terms of perturbative quantum gravity, the metric is expanded about some arbitrary classical background $g=\langle g \rangle+ \epsilon_g$ where the equilibrated fluctuations $\epsilon_g$ are governed by finite-temperature quantum field theory.  The $\epsilon_g$ fluctuation's back-reaction against $\langle g \rangle$ dynamically drives $\langle g \rangle$ which is `on-shell' (obeys expectation-valued Einstein equations) when the $\epsilon_g$ fluctuation entropy is maximum.   
\end{sloppypar}
\begin{sloppypar}
To identify the correct quantum description of fluctuations about this background, we utilize the quantum gravity path integral whose measure $d[g]$ can be written as the product of classical backgrounds $d[\langle g \rangle]$ and the zero-point quantum fluctuations $d[\epsilon_g]$ about these classical backgrounds:
\begin{equation}\label{fullpath1}
Z=\int d[g]e^{-I(g)}=\int d[\langle g \rangle]d[\epsilon_g]e^{-I(\langle g \rangle, \,\epsilon_g)}
\end{equation}
Because we have the many-body probability distributions for $\epsilon_g$ at our disposal, we can calculate their entropy $S$ (which is a functional of the background) and then maximize this entropy by performing an Euler-Lagrange variation of $S$ with respect to the background.  This picks out an on-shell (entropy-maximizing) family of backgrounds $\langle g \rangle$ which obey the expectation-valued Einstein equations.  
\end{sloppypar}
\begin{sloppypar}
Physically, the many-body equilibrium of the $\epsilon_g$ zero-point fluctuations restricts the path integration over $d[g]$ to a sub-manifold of its phase-space in which the $\epsilon_g$  fluctuation entropy is maximum.  The background $\langle g \rangle$ is `fixed' via entropy maximization, the measure $d[\langle g \rangle]$ is restricted to this background, and the remaining path integration over $d[\epsilon_g]$ presents a finite-temperature quantum description of the fluctuations about this background\footnote{The background has coordinates $\langle x \rangle$, a metric $\langle g \rangle$, and a Ricci scalar $\langle R \rangle$ which equals the expectation of the quantum Ricci scalar $R$.}:
\begin{equation}
Z \rightarrow Z_{\langle g \rangle}=\int d[\epsilon_g]e^{-\beta I_{\langle g \rangle}(\epsilon_g)}
\end{equation} 
\begin{equation}
I \rightarrow I_{\langle g \rangle}=\int e^{-L_G^2 \langle R \rangle}\bigl(R-\langle R \rangle\bigr)^2 d^D\!\langle x \rangle \sqrt{\negthinspace\lvert \langle g \rangle \rvert}
\end{equation}
\begin{equation}
\beta=\ell_0^4 
\end{equation}
This path integral, action, and temperature parameter are worked out explicitly in section \ref{intpic} where it is noted that this quantum formulation is manifestly background-independent. 
\end{sloppypar}
\begin{sloppypar}
It is important to note that the $R$-squared action $I_{\langle g \rangle}$ presents, in a certain sense, an \textsl{interaction picture} of quantum gravity.  In this picture we are `moving with' the classical background $\langle g \rangle$ in Hilbert space, and the action $I_{\langle g \rangle}$ encodes the quantum description of fluctuations about $\langle g \rangle$.  The `back-reaction' of the quantum fluctuations $\epsilon_g$ against $\langle g \rangle$ was fully accounted for when we maximized their entropy which put $\langle g \rangle$ on-shell.  The dynamics of the classical background $\langle g \rangle$ are not determined from an Euler-Lagrange variation of $I_{\langle g \rangle}$ with respect to ${\langle g \rangle}$.  Rather, it is the \emph{entropy}\footnote{The entropy of the quantum Riemann tensor's zero-point fluctuations.} $S$ of $Z_{ \langle g \rangle}$ and the Euler-Lagrange variation of $S$ with respect to ${\langle g \rangle}$ which determines the background.  
\end{sloppypar}
\begin{sloppypar}
A similar situation exists in the thermodynamics of a gas:  Entropy maximization determines velocity distribution which determines the expected velocity field $\langle v \rangle$, while the individual molecules obey a finite-temperature action about $\langle v \rangle$.  The entropy of the gas is maximized via local energy exchange, while the entropy of quantum spacetime is maximized via local curvature exchange.  
\end{sloppypar}
\begin{sloppypar}
Within this interaction picture, the fields required to describe quantum gravity have been transformed into the `ordinary' quantum field $\mathcal{R}\equiv R-\langle R \rangle$ which propagates on the `rigid' or `fixed' background $\langle g \rangle$ with action $I_{\langle g \rangle}$ and propagator $\frac{1}{k^4+Ak^2}\,$.   We quantize the fluctuations about $\langle g \rangle$, but we do not quantize the \textsl{expected} geometry $\langle g \rangle$.  The finite-temperature path integral $Z_{ \langle g \rangle}$, along with its action $I_{\langle g \rangle}$, provide us with the correlators and propagators for the quantum fluctuations about any particular on-shell background $\langle g \rangle$.     
\end{sloppypar}
\begin{sloppypar}
Throughout the paper several boost-invariant lengths appear as a consequence of statistical physics \cite{DSR1, DSR2}.  There is some discussion in section \ref{intpic} as to how we might couple matter fields to this entropy-driven background and the implications this might have on both the dark-matter problem and the zero-point energy cosmological constant problem.  Also mentioned in section \ref{intpic} is the ability of vacuum's curvature fluctuations to transport an evaporating black hole's entropy which provides a mechanism to account for black hole information loss \cite{hawkingradiation, BHinfoloss}.   
\end{sloppypar}
\section{Quantum expectation and zero-point fluctuations}
\begin{sloppypar}
At Planck scale, quantum spacetime presumably has a large number of available states.  If the spectrum $\vartheta$ and probability distribution $P_{\vartheta}$ of these states were known, their entropy $S=-\sum P_{\vartheta} \ln P_{\vartheta}$ would be maximum because of the following arguments.     
\end{sloppypar}
\begin{sloppypar}
Suppose we wanted to examine the ground state of the quantum field $g=\eta+\epsilon_g$ at the point $q$ where $\epsilon_g$ are the zero-point fluctuations about the Minkowski background.  If we had at our disposal the eigenvectors $\vert R_A \rangle$ of the curvature operator $\hat{R}_{\mu \nu \rho \sigma}\,$, then we immediately recognize that the curvature state at $q$ must be a \textsl{mixed state} because we are examining the zero-point fluctuations.  The expectation of curvature at $q$ is then:
\begin{equation}\label{state}
\langle \hat{R}_{\mu \nu \rho \sigma} \rangle = \sum w_A\langle R_A  \vert \hat{R}_{\mu \nu \rho \sigma} \vert R_A \rangle=0
\end{equation}
where $w_A \in \mathbb{R}$ and $\sum w_A =1$.
\end{sloppypar}
\begin{sloppypar}
Because $\epsilon_g$ is a spin two quantum field, the mixed curvature state must be an isotropic sum over helicity states (helicity states correspond with anisotropic curvature fluctuations) otherwise vacuum would be anisotropic at the quantum level.  Additionally, if the quantum dynamics of $\hat{R}_{\mu \nu \rho \sigma}$ prevent us from simultaneously knowing all the components of $\hat{R}_{\mu \nu \rho \sigma}\,$, then the mixed curvature state must be isotropic with respect to these `component' states as well.  Alternatively, expectation could be defined as an ensemble average over a single component of curvature (during which the other components are not measured) and then repeating the ensemble averaging, component by component, over each physically independent  (but not necessarily statistically independent) component of curvature.    
\end{sloppypar}
\begin{sloppypar}
In addition to expectation, each individual component of $\hat{R}_{\mu \nu \rho \sigma}$ obeys a probability distribution at the point $q$.  Because the fluctuations must appear the same to all inertial observers, these distributions must be invariant under the Poincare group.  If they depended explicitly on position and/or orientation, vacuum would fail to be homogeneous and/or isotropic at the quantum level.  If they were boost-dependent, an ether and prefered rest frame would exist.  The lack of explicit positional dependence rules out time-dependent coefficients ($\partial_t w_A=0$) in the mixed curvature state which rules out diffusion in Hilbert space.  Since the zero-point curvature fluctuations are interacting with each other (gravitation \emph{self-interacts}) I conclude from their absence of diffusion in Hilbert space that they are in many-body equilibrium with each other.  It should be noted that arguments against this conclusion which rely on the homogeneity and isotropy of the background $\eta$ are circular:  The homogeneity and isotropy of $\eta$ (which is an artifice since $\eta$ is not the actual geometry, $g=\eta+\epsilon_g$ is the actual geometry) exists because $g$ is homogeneous and isotropic, and this property of $g$ exists because the fluctuations $\epsilon_g$ are homogeneous and isotropic.     
\end{sloppypar}
\begin{sloppypar}
I now want to generalize these ideas to an arbitrary quantum spacetime which begins by constructing its background from the expectations $\langle  \hat{R}_{\mu \nu \rho \sigma} \rangle$, $\langle  \hat{R}_{\mu \nu} \rangle$, and $\langle \hat{R} \rangle$.  This requires identifying the point $\langle x \rangle $ of some classical background with the point $q\langle x \rangle$ of quantum spacetime, and statistical physics will later show this assumption of a local diffeomorphism between the classical and quantum geometries to be valid.  The background $\langle g \rangle_{\mu\nu}$ has coordinates $\langle x \rangle_{\mu}$ and curvature $\langle  R \rangle_{\mu \nu \rho \sigma\,}$, where brackets inside the indices serve as labels indicating `classical background' and do not indicate the expectation operator.  We are free to choose the background, so we choose one with curvature at $\langle x \rangle$ which numerically equals quantum spacetime's expected curvature at $q\langle x \rangle$:
\begin{subequations}\label{background}
\begin{gather}
\langle R \rangle_{\mu \nu \rho \sigma} = \langle \hat{R}_{\mu \nu \rho \sigma} \rangle \label{a}\\ 
\langle R \rangle_{\mu \nu} = \langle \hat{R}_{\mu \nu} \rangle \label{b}\\
\langle R \rangle = \langle \hat{R} \rangle \label{c} 
\end{gather}
\end{subequations} 
By repeating this process throughout the neighborhoods $\langle U \rangle$ of $\langle x \rangle $ and $W=q\langle U \rangle$, the background becomes the \emph{locally curvature averaged} version of the quantum geometry within $\langle U \rangle$.  
\end{sloppypar}
\begin{sloppypar}
The quantum Ricci scalar's zero-point fluctuations about the background, $\mathcal{R} = R-\langle R \rangle$, obey some probability distribution $P_\mathcal{R}$ within $\langle U \rangle$ on the background.  Within `macroscopically small' neighborhoods (such as an Einstein freely-falling cabin which is small compared to large-scale structure, yet large compared to small-scale structure) the previous arguments of homogeneity and isotropy still apply.  The zero-point curvature fluctuations are in equilibrium with each other on the general background, and this equilibrium is expressed as:
\begin{equation}\label{equilibrium}
\partial_{\mu} P_{\mathcal{R}}=0
\end{equation}
where $\partial_{\mu}$ is differentiation with respect to the background $\langle x \rangle_{\mu}$.  While the background $\langle g \rangle$ can and will dynamically evolve, it is the zero-point fluctuations $\mathcal{R}$ about the background that are in many-body equilibrium with each other.  To emphasize the many-body nature of these fluctuations, a typical atom's volume divided by Planck's volume is about $4 \times 10^{73}$.  Because gravitation self-interacts, this is an order of magnitude estimate of the number of gravitationally-coupled systems within this volume.  Each of these systems have their own Planck-scale curvature degrees of freedom, and it is these degrees of freedom which obey the laws of statistical physics.  Because $\partial_{\mu} P_{\mathcal{R}}=0$ the curvature fluctuations do not diffuse, and this lack of diffusion is an example of decoherence in quantum gravity \cite{decohere1, decohere}.  
\end{sloppypar}
\begin{sloppypar}
The distribution $P_\mathcal{R}$ must be entropy maximizing.  If it was not, the $\mathcal{R}$ fluctuations would, like the fluctuations within all self-contained many-body systems, redistribute themselves into the most likely phase-space configuration.  While this was occuring, the variance of $\mathcal{R}$ (its canonical temperature) would depend explicitly on $\langle x \rangle _{\mu}$ which violates local homogeneity.  Currents would flow (just as heat flows from hot to cold in ordinary matter) until an entropy maximizing $P_{\mathcal{R}}$ and local homogeneity were established.  Quantum spacetime's Planck-scale spectrum of available states $\vartheta$ must then (by the same arguments as $\mathcal{R}$ fluctuations) be distributed with an entropy maximizing $P_{\vartheta}$ which proves the assertion at the beginning of this section. 
\end{sloppypar}
\section{The quantum Riemann tensor's entropy}\label{manybody}
\begin{sloppypar}
Denote the $N=D^2 (D^2-1)/12$ physically independent components of $R_{\mu \nu \rho \sigma}$ by $R_A$.  If we measured a single component ($R_1$ for example) at the point $\langle x \rangle$ without regard to the other component's values, then $R_1$ would obey some probability distribution $P_1$.  If we wanted to \textsl{simultaneously} measure two (or more) components in a compound event, it must be assumed that the $R_A$ are in general statistically dependent variables (just as with spin components which cannot be simultaneously known) therefore the compound event outcomes obey conditional probabilities.  In other words, the entropy of the quantum Riemann tensor is a difficult sum over its not-yet-known conditional probabilities.  
\end{sloppypar}
\begin{sloppypar}
We can, however, perform a transformation on the $R_A$ after which one component will be statistically independent from the others.  To see this, consider the unrestricted phase-space\footnote{The set of all possible values.} of $R_{\mu \nu \rho \sigma}$ at $\langle x \rangle$.  This phase-space $\Omega_{Riem}$ is spanned by orthonormal $e_A$ and has coordinates $R_A$.  It is interesting to note that from the pole of a normal coordinate system, the $R_A$ are simply the physically independent components of $\frac{1}{2} \delta_{ij}^{ab} \delta_{kl}^{cd} \partial_{a} \partial_{c} g_{bd}$.  Since $R^{\sigma \rho}_{\quad\sigma \rho}=R$ induces a linear map from $R_A$ to $R$ (it defines an $N-1$ dimensional hyperplane of constant $R$ within $\Omega_{Riem}$) we can always perform an $SO(N)$ transformation on the basis $e_A$:
\begin{equation}\label{trans}
R_{\bar{A}}=M_{\bar{A}B} R_B
\end{equation}
such that one of the new coordinates, say $R_{\bar{1}}$, is the Ricci scalar $R$.  Subtracting the background's Ricci scalar $\langle R \rangle $ from this coordinate, $\mathcal{R}=R-\langle R \rangle$, and then dropping the overbars on the remaining coordinates for clarity, the phase-space of all possible curvature fluctuations about the background at $\langle x \rangle$ becomes (where $A$ now runs from $2$ to $N$):
\begin{equation}\label{omega}
\Omega_{Riem}=\Omega_{\mathcal{R}} \times \Omega_{R_A}
\end{equation}
The phase-space $\Omega_{\mathcal{R}}$ represents only active diffeomorphisms, and it will be proven shortly that its coordinate $\mathcal{R}$ is statistically independent from the remaining coordinates $R_A$.  The phase-space $\Omega_{R_A}$ represents passive as well as active diffeomorphisms, and its coordinates $R_A$ will in generally be statistically dependent on one another.  The Gibbs entropy of the quantum Riemann tensor at $\langle x \rangle$ is then: 
\begin{equation}\label{rte}
S_{\langle x \rangle}^{Riem}=-\sum P_{\mathcal{R}} \ln P_{\mathcal{R}} + S({R_A}) 
\end{equation}
We find the explicit form of $P_{\mathcal{R}}$ by examining the grand-canonical fluctuations of a quantum field which creates and destroys quanta of vacuum, and with this $P_{\mathcal{R}}$ the statistical independence of $\mathcal{R}$ from $R_A$ becomes manifest.  
\end{sloppypar}
\begin{sloppypar}
Since the universe is expanding, vacuum is being created.  If we accept that vacuum is created by the operators of some quantum field, then vacuum is created locally in discrete units.  The only way to reconcile this discreteness with local Lorentz invariance is with discrete spacetime \emph{events} whose local relative positions obey complete spatial-temporal randomness \cite{burton}, therefore the operators act in events (with local relative positions that obey complete spatial-temporal randomness) which create and destroy quanta of spacetime volume.  Since these operators do not exist on a background (they locally create and destroy spacetime volume itself) they are by definition incompressible.  Because they have constant density, their grand-canonical fluctuations $\frac{N-\langle N \rangle}{\langle N \rangle}$ appear as local volume fluctuations $\frac{V-\langle V \rangle}{\langle V \rangle}$, and these local volume fluctuations cause local curvature fluctuations that are in many-body equilibrium with each other.  
\end{sloppypar}
\begin{sloppypar}
Each small volume element of spacetime is acting as a grand-canonical fluctuating system which is in fixed-density chemical equilibrium with the rest of causally-connected spacetime (the reservoir).  The local volume fluctuations, as well as their local curvature fluctuations, can be probed using the volume deviation (relative to the background) of a geodesic ball:
\begin{equation}\label{del}
\delta_V \equiv\frac{V-\langle V \rangle}{\langle V \rangle}= \frac{-1}{6(D+2)}\mathcal{R}\,\varepsilon^2+O(\varepsilon^3)
\end{equation}      
where $\varepsilon$ is the radius of a geodesic ball, $\langle V \rangle$ is the expected (the background's) ball volume at radius $\varepsilon$, $V$ is the actual volume, $\mathcal{R}= R-\langle R \rangle$, the background's Ricci scalar is $\langle R \rangle$, and the quantum Ricci scalar is $R$.  In the limit of high operator (or volume quanta) density, $\delta_{V}$ becomes a continuous Gaussian random variable.  In this limit, by the continuity theorem of transforms of normal distributions, $\mathcal{R}$ is also normally distributed:
\begin{equation}\label{dev}
\frac{\partial^2 \delta_{V}}{\partial \varepsilon^2 }\Bigr\vert_{\varepsilon=0}=\frac{-1}{3(D+2)} \mathcal{R} 
\end{equation}
\begin{equation}\label{PofRcurved}
P_{\mathcal{R}}d\mathcal{R}=\frac{1}{\sqrt{\pi}\,\sigma_{\mathcal{R}}} e^{-\mathcal{R}^2 / \sigma^2_{\!\mathcal{R}}}\,d\mathcal{R}
\end{equation} 
Since $P_{\mathcal{R}}$ does not depend on the phase-space variables $R_A$, the claim of statistical independence  is proven.  
\end{sloppypar}
\begin{sloppypar}
The curvature variance $\sigma_{\mathcal{R}}^2$ can also be written in terms of a \textsl{correlation length} or \textsl{temperature}:
\begin{equation}\label{coorlength}
\sigma^2_{\mathcal{R}} \equiv 1 / \ell^4 \propto T  
\end{equation}
Because $P_{\mathcal{R}}$ has a finite temperature (or variance or correlation length) quantum geometry must become correlated at short distance.  The absence of correlations at any scale corresponds with an infinite temperature system.  Many-body interactions keep `large' curvature fluctuations Boltzmann suppressed which keeps the small-scale geometry locally smooth.  This provides the justification for the local diffeomorphism between quantum and classical geometry, and permits the use of tools such as normal coordinate expansions and volume deviation formulas.  This small-scale smoothness stands in sharp contrast with the uncertainty-based picture of ever-increasing fluctuations as scale decreases \cite{kiefer} which does not take into account many-body interaction equilibrium.  The statistical field theory of membranes and surfaces exhibits similar behavior where membrane normal vectors become correlated \cite{membranecorrelation} at short distance:  
\begin{equation}\label{normals}
\langle \hat{n}(\mathbf{x})\cdot \hat{n}(0) \rangle \propto e^{-x/\xi}
\end{equation}
\end{sloppypar}
\begin{sloppypar}
The quantum Riemann tensor's zero-point fluctuations `back-react' against the background because their variance, and hence their entropy $S^{Riem}_{\langle x \rangle} \sim \ln(\sqrt{\pi e}\,\sigma_{\mathcal{R}})$, depends on the background:
\begin{equation}\label{sigmaR}
\sigma_{\mathcal{R}}^2=\sigma_0^2 e^{L_G^2 \langle R \rangle}
\end{equation}
Because $S^{Riem}_{\langle x \rangle}$ depends on the background's configuration $\langle R \rangle$, maximizing the total fluctuation entropy requires summing $S^{Riem}_{\langle x \rangle} \sim \langle R \rangle$ over the background's volume and then varying the background's configuration.  To prove equation \eqref{sigmaR}, vary the background's curvature while holding a geodesic ball at constant expected volume $\langle V \rangle=\langle V_1 \rangle=\langle V_2 \rangle$:
\begin{subequations}\label{deformvolumes}
\begin{align}
\delta_{V_1} = \frac{V_1-\langle V_1 \rangle}{\langle V_1 \rangle}=\frac{-1}{6(D+2)}\mathcal{R}_1 \,\varepsilon^2_1+O(\varepsilon^3_1)\label{a}\\
\delta_{V_2} = \frac{V_2-\langle V_2 \rangle}{\langle V_2 \rangle}=\frac{-1}{6(D+2)}\mathcal{R}_2\,\varepsilon^2_2+O(\varepsilon^3_2)\label{b} 
\end{align}
\end{subequations}
Since $\langle V_1 \rangle=\langle V_2 \rangle$ in equations \eqref{deformvolumes}, $\delta_{V_1}$ and $\delta_{V_2}$ obey the same probability distribution.  Since we are working at constant expected volume, if $\langle R_1 \rangle \neq \langle R_2 \rangle$ then $\varepsilon_1 \neq \varepsilon_2$.  This implies the variance of $\mathcal{R}_1$ differs from the variance of $\mathcal{R}_2$, and this proves the variance of $\mathcal{R}$ depends on the background.
\end{sloppypar}
\begin{sloppypar}
Next, vary the background's curvature while holding a geodesic ball at constant radius $\varepsilon =\varepsilon_1 = \varepsilon_2$.  Since any system in chemical equilibrium with a reservoir has grand-canonical fluctuations $\delta_{N} = \frac{N-\langle N \rangle}{\langle N \rangle}$ whose variance $\sigma^2_{N}$ increases with decreasing system size $\langle N \rangle$, the variance of $\delta_{V}$ will also increase with decreasing system size $\langle V \rangle$.  Because we are working at constant ball radius, if $\langle R_2 \rangle > \langle R_1 \rangle$ in equations \eqref{deformvolumes}, volume deviation results in $\langle V_2 \rangle < \langle V_1 \rangle$.  Therefore the variance of $\delta_{V_2}$ will be larger than the variance of $\delta_{V_1}$, and this implies the variance of $\mathcal{R}$ increases with increasing $\langle R \rangle$.  
\end{sloppypar}
\begin{sloppypar}
To exhibit the explicit dependence of $\sigma^2_{\mathcal{R}}$ on $\langle R \rangle$, consider the zero-point fluctuations of the quantum Ricci scalar on a background which has $\langle R \rangle =0$.  If the number of available fluctuation states at a point on the background is $\Omega$, then (using $\rho=$ number of curvature systems per background volume) the number of available states in a small volume $\langle dV \rangle$ will be $\Omega^{\rho \langle dV \rangle}$.  If we perform a curvature variation $\langle \delta R \rangle$ on this background while holding the volume's size $\langle dV \rangle$ fixed, the number of available states becomes: 
\begin{equation}\label{states}
\Omega^{\rho \langle dV \rangle (1+ \delta f)}
\end{equation}
for some function $f\langle R \rangle$, and the entropy within $\langle dV \rangle$ becomes:
\begin{equation}\label{volentropy}
dS = \rho \langle dV \rangle \left( 1+ \delta f \right) \ln \Omega  
\end{equation} 
Because the quantum Ricci scalar's zero-point fluctuations are in equilibrium with each other (meaning $\langle dV \rangle$ is an extensive thermodynamic variable) their entropy within $\langle dV \rangle$ can also be written, using $S_{\langle x \rangle}^{Ricci}=-\sum P_{\mathcal{R}} \ln P_{\mathcal{R}}$, as: 
\begin{equation}\label{volentropy2}
dS=\rho \langle dV \rangle \ln (\sqrt{\pi e}\,\sigma_{\mathcal{R}})
\end{equation}
Comparing equations \eqref{volentropy} and \eqref{volentropy2}, the quantum Ricci scalar's variance depends on background curvature variations as: 
\begin{equation}\label{sigmaR3}
\sigma_{\mathcal{R}}^2=\sigma^2_0 e^{\delta f}
\end{equation} 
Lastly, since two derivatives of volume deviation $\frac{\partial^2 \delta_{V}}{\partial \varepsilon^2 }\Bigr\vert_{\varepsilon=0}\negthickspace=\!\frac{-1}{3(D+2)} \mathcal{R}\,$ evaluated at $\varepsilon=0$ excludes fluctuations from higher-order terms, the variance $\sigma_{\mathcal{R}}^2\,$ cannot depend on higher-order background derivatives.  The function $f$ is therefore linear in $\langle R \rangle$, and the variance of the quantum Ricci scalar  is:  
\begin{equation}\label{sigmaR2}
\sigma_{\mathcal{R}}^2=\sigma^2_0 e^{L_G^2 \langle R \rangle}
\end{equation} 
were the parameter $L_G$ is a universal constant.  The factor $e^{L_G^2 \langle \delta R \rangle}$ behaves as an element of a Lie group which acts on the quantum Ricci scalar's number of available states.   
\end{sloppypar}
\begin{sloppypar}
The quantum Riemann tensor's entropy at the point $\langle x \rangle$, equation \eqref{rte}, receives contributions from $\mathcal{R}$ and the phase-space variables $R_A$.  To prove $S(R_A)=0$, we appeal to a condensed-matter system with similar fluctuations.  Suppose we had a two-dimensional, temperature-sensitive membrane which locally expands or contracts according to:
\begin{equation}\label{membrane}
\frac{A-\langle A \rangle}{\langle A \rangle}=k\bigl( T-\langle T \rangle \bigr)
\end{equation}
As long as the membrane's temperature is maintained everywhere at $\langle T \rangle$, an initially flat membrane remains flat.  But if $T$ has spatial variations which are distributed normally (and with a position-independent variance) about $\langle T \rangle$, the local expansion and contraction variations cause Ricci scalar variations which are distributed normally (and with a position-independent variance) about $\langle R \rangle =0$.  Since the local expansion and contraction variations uniquely determine the membrane's configuration $\, g_{ij}$ (via 2nd order field equations with temperature gradients as sources) the Riemann tensor's location $(\mathcal{R}, R_A)$ within its phase-space is uniquely determined by $T$ at each point on the membrane.  Therefore we cannot treat both $\mathcal{R}$ and $R_A$ as fluctuating variables which would over-count the entropy of the local expansions and contractions.  Since the source of these fluctuations is the scalar $T$ whose spatial variations are directly related (via volume deviation) to $\mathcal{R}$, the membrane's entropy resides in $\mathcal{R}$.  
\end{sloppypar}
\begin{sloppypar}
A quantum field which locally creates and destroys spacetime volume is a background-free version of the temperature-sensitive membrane.  Grand-canonical volume fluctuations correspond with the membrane's area fluctuations, and the volume fluctuation entropy resides in $\mathcal{R}$.  This proves $S(R_A)=0$.      
\end{sloppypar}
\section{The background's equations of motion}\label{eom}
\begin{sloppypar}
Since $\sum P_{\mathcal{R}} \ln P_{\mathcal{R}}$ sums over a continuum, we need the quantum Ricci scalar's phase-space density of states $\rho_R$.  This density of states is constant for the following reasons:  $\Omega_{Riem}$ is a linear vector space with basis $e_A$ and coordinates $R_A$, the transformation $R_{\bar{A}}=M_{\bar{A}B} R_B$ was $SO(N)$, the volume of $\Omega_{R_A}$ (constant-$R$ hyperplanes) is independent of $R$, and $S(R_A)=0$.  A convenient way to introduce $\rho_R\equiv \ell_{R}^2$ is by multiplying and dividing throughout $P_{\mathcal{R}}$ by $\ell_{R}^2$, changing variables to the dimensionless $\tilde{\mathcal{R}}=\ell_{R}^2 \mathcal{R}$, and then dropping the tilde for clarity:
\begin{equation}\label{PofRcurved2}
P_{\mathcal{R}}d\mathcal{R}=\frac{1}{\sqrt{\pi}\,\sigma} e^{-\mathcal{R}^2 /\sigma^2}\,d\mathcal{R}
\end{equation}
where the variance is now $\sigma^2 = \ell_{R}^4  \sigma^2_{\mathcal{R}}$.  From equation \eqref{volentropy2}, using $\sigma^2 =\ell_R^4 \sigma_0^2 e^{L_G^2 \langle R \rangle}$ and writing $\sigma^2_0$ as a correlation length $\ell_0^4 \equiv 1/\sigma^2_0$, the entropy in a small volume $\langle dV \rangle$ is:
\begin{equation}\label{flucentropy}
dS=\rho \langle dV \rangle \left[\ln\left(\frac{\sqrt{\pi e}\,\ell^2_{R}}{\ell_0^2}\right)+\frac{1}{2}L_G^2\langle R \rangle\right]
\end{equation}
Integrating over the background using $\langle dV \rangle \equiv d^D\!\langle x \rangle \sqrt{\negthinspace\lvert \langle g \rangle \rvert}$ to get the total entropy $S$, and then bringing the coefficient of $\langle R \rangle$ to $1$ yields: 
\begin{equation}\label{varS}
\frac{2S}{\rho L^2_G}=\int\negthickspace d^D\!\langle x \rangle \sqrt{\negthinspace\lvert \langle g \rangle \rvert}\,\Bigl( \langle R \rangle+2\Lambda \Bigr) =0
\end{equation}
\begin{equation}\label{lamda}
\Lambda=\frac{1}{L_G^2}\ln\left(\frac{\sqrt{\pi e}\,\ell_{R}^2}{\ell_0^2}\right)
\end{equation}
The Euler-Lagrange variation of $S$ with respect to the background $\langle g \rangle$ yields the on-shell, or entropy-maximizing, family of backgrounds:
\begin{equation}\label{MFE}
\langle G \rangle_{\mu\nu} -\Lambda \langle g \rangle_{\mu\nu}=0
\end{equation}
which are the expectation-valued Einstein equations.
\end{sloppypar}
\begin{sloppypar}
The quantum Riemann tensor's zero-point fluctuations create inherent length scales.  The correlation length $\ell_0$ is a topological invariant under diffeomorphisms while $\ell^4 = \ell_0^4 e^{-L_G^2 \langle R \rangle}$ is invariant under boosts.  The quantum Ricci scalar's phase-space density of states $\ell_{R}^2$ and the parameter $L_G $ are also topological invariants under diffeomorphisms.  All of these lengths are boost-invariant because of statistical physics and do not require a modification of special relativity to explain their constancy.    
\end{sloppypar}
\section{Entropy driven expansion}\label{expansion}
\begin{sloppypar}
The cosmological constant $\Lambda=\frac{1}{L_G^2}\ln\left(\frac{\sqrt{\pi e}\,\ell_{R}^2}{\ell_0^2}\right)$ effectively depends on one parameter $\ell_{R}^2 / \ell_0^2 \sim \sqrt{T}$ which plays the role of a curvature fluctuation temperature, and the constant $L_G$ can in principle be derived from probability theory.  The expansion caused by $\Lambda$ has several interpretations.  It can be viewed as the consequence of curvature fluctuation pressure, or the back-reaction to the zero-point curvature fluctuations.  The expansion can also be viewed as the tendency, inherent in all self-contained many-body systems, to configure at maximum entropy:  A quantum field creates (and destroys) vacuum, more vacuum means more entropy, therefore the universe will tend to expand.  Another viewpoint comes from the condensed-matter fluid membrane.  These fluid membranes (whose actions are identical, apart from bending rigidity terms, to the Einstein-Hilbert action) have their own `cosmological constant' \cite{membranelambda} which quantifies the chemical potential (per unit area) of membrane creation. 
\end{sloppypar}
\section{Finite-temperature R-squared quantum gravity}\label{intpic}
\begin{sloppypar}
To identify the correct quantum description of fluctuations about the background $\langle g \rangle$, we construct the finite temperature, quantum gravity path integral from the curvature fluctuation probability distributions.  
\end{sloppypar}
\begin{sloppypar}
We begin with the Gaussian $P_{\mathcal{R}}$ of equation \eqref{PofRcurved} and its variance $\sigma^2_{\mathcal{R}} = \sigma_0^2 e^{L_G^2 \langle R \rangle}$, where it is recalled that $\mathcal{R}=R-\langle R \rangle$ and that $d \mathcal{R}=dR$ since the background has already been fixed on-shell:
\begin{equation}\label{PofRcurved3}
P_{\mathcal{R}}=\frac{e^{-\mathcal{R}^2 / \sigma^2_{\!\mathcal{R}}}}{\int\!d\mathcal{R}\,e^{-\mathcal{R}^2 / \sigma^2_{\!\mathcal{R}}}} 
\end{equation} 
From the general property of partition functions $P_{\mathcal{H}}=e^{-\beta \mathcal{H}}/Z$, the partition function for curvature fluctuations at a point on the background is:
\begin{equation}
Z^{point}_{\langle g \rangle}=\int d\mathcal{R}\,e^{-\beta\mathcal{G}\mathcal{R}^2} 
\end{equation}
where $\beta\mathcal{G}=1/\sigma^2_{\mathcal{R}}$ and $\mathcal{G}=e^{-L_G^2 \langle R \rangle}$.  From the temperature parameter $\beta=1/\sigma^2_0 =\ell^4_0\,$ we see that an absolute zero temperature $T=1/\beta=1/\ell^4_0$ corresponds with an infinite correlation length $\ell_0\,$, while an infinite temperature $T$ corresponds with a zero correlation length $\ell_0$. 
\end{sloppypar}
\begin{sloppypar}
Next, we divide the background's total volume $V$ into small cells\footnote{Small means quantum geometry within $\Delta V$ is highly correlated which means $\mathcal{R}=R-\langle R \rangle$ does not vary appreciably within $\Delta V$.} of equal volume $\Delta V$ where the $j$th cell has volume $\bigl(\Delta V\bigr)_j \equiv \bigl(d^D\!\langle x \rangle \sqrt{\negthinspace\lvert \langle g \rangle \rvert}\,\bigr)_j\,$ and then write down the partition function for one cell:
\begin{equation}
Z^{cell}_{\langle g \rangle}=\int d\mathcal{R}\,e^{-\beta\mathcal{G}\mathcal{R}^2 (\frac{\Delta V}{\Delta V})}=\sqrt{\Delta V}\int d\mathcal{R}\,e^{-\beta\mathcal{G}\mathcal{R}^2 \Delta V}
\end{equation}
Taking the product of all cell partition functions gives us the partition function for the total volume $V$:
\begin{equation}
Z_{\langle g \rangle}= (\Delta V)^{(N/2)}\!\int \prod_{j=1}^{N}  d\mathcal{R}_j \, e^{-\beta \left(\mathcal{G}\mathcal{R}^2 \Delta V\right)_j}
\end{equation}
where $N=V/ \Delta V$.  Taking the limit $\Delta V \rightarrow 0$ gives us the finite-temperature path integral:
\begin{equation}
Z_{\langle g \rangle}=\frac{1}{\mathcal{N}} \int d[\mathcal{R}]e^{-\beta I_{\langle g \rangle}}
\end{equation}
\begin{equation}
I_{\langle g \rangle}=\int e^{-L_G^2 \langle R \rangle}\Bigl(R-\langle R \rangle\Bigr)^2 d^D\!\langle x \rangle \sqrt{\negthinspace\lvert \langle g \rangle \rvert}
\end{equation}
\begin{equation}
\beta=\ell_0^4 
\end{equation}
where $\mathcal{N}=\rho^{\frac{\rho V}{2}}$ and $\rho = N / V$.  Writing out the full path integral over all backgrounds, equation \eqref{fullpath1}, the measure $d[\epsilon_g]$ over metric fluctuations has become a measure $d[\mathcal{R}]$ over curvature fluctuations:  
\begin{equation}\label{fullpath2}
Z=\int d[g]e^{-I(g)}=\int d[\langle g \rangle]d[\mathcal{R}]e^{-I(\langle g \rangle, \,\mathcal{R})}
\end{equation}
\end{sloppypar}
\begin{sloppypar}
As mentioned in the introduction, we do not determine the background via an Euler-Lagrange variation of $I_{\langle g \rangle}$ with respect to $\langle g \rangle$.  The background has already been determined via entropy maximization of the quantum Riemann tensor's fluctuations about the background.  The quantum field $\mathcal{R}=R-\langle R \rangle$ is now an `ordinary' quantum field which propagates on the `rigid' or `fixed' background $\langle g \rangle$ with action $I_{\langle g \rangle}$ and propagator $\frac{1}{k^4+Ak^2}\,$.  Since the action's quadratic term $\bigl(R- \langle R \rangle\bigr)^{\!2}$ is invariant under exchange of the $\langle\, \rangle$ operator, and the field $\langle R \rangle$ initially has no preferred value (it is dynamically determined via entropy maximization) this quantum formulation is manifestly background independent.      
\end{sloppypar}
\begin{sloppypar}
A consistent method of coupling matter fields to the entropy-driven background $\langle g \rangle$ would treat the zero-point fluctuations of both matter fields and geometry on an equal footing.  For an arbitrary background $\langle g \rangle$, we could expand both geometry and matter fields about their zero-point fluctuations $g=\langle g \rangle+\epsilon_g$ and $\Psi=\langle \Psi \rangle+ \epsilon_{\Psi}$, find the many-body probability distributions for $\epsilon_g$ and $\epsilon_{\Psi}$, and then compute the total zero-point fluctuation entropy $S$.  The Euler-Lagrange variation of $S$ with respect to $\langle g \rangle$ should give us the on-shell background $\langle g \rangle$ which is coupled to $\langle \Psi \rangle$, and the Euler-Lagrange variation of $S$ with respect to $\langle \Psi \rangle$ should give us the `on-shell' $\langle \Psi \rangle$ which is coupled to $\langle g \rangle$.  The correlators and propagators of both $\epsilon_g$ and $\epsilon_{\Psi}$ can then be constructed from the finite-temperature path integral:   
\begin{equation}
Z_{\langle g \rangle \langle \Psi \rangle}=\int d[\epsilon_g]d[\epsilon_{\Psi}]e^{-\beta I_{\langle g \rangle \langle \Psi \rangle}}
\end{equation} 
\end{sloppypar}
\begin{sloppypar}
Since matter's entropy is insensitive to what we call the zero-point of energy, the resulting equations of motion for $\langle g \rangle$ should be insensitive to shifts in the zero-point of energy for $\langle \Psi \rangle$, therefore this procedure should ameliorate the cosmological constant's zero-point energy problem \cite{weinberg}.  Additionally, the entropy of gravitating matter sources should lead to source correction terms which would exhibit `dark-matter' behavior.                 
\end{sloppypar}
\begin{sloppypar}
While black hole thermodynamic relations are well known \cite{bhentropy, pope}, their microscopic origins are less well known.  Knowing the quantum Riemann tensor's entropy, and knowing how to couple this to matter fields, should provide some insight.  Since vacuum's curvature fluctuations have entropy $dS \propto \sqrt{\lvert \langle g \rangle \rvert}d^D\langle x \rangle \bigl(\langle R \rangle + 2\Lambda \bigr)$, they have the ability to transport an evaporating black hole's entropy which provides a mechanism to account for black hole information loss.     
\end{sloppypar}
\section{Conclusion}\label{conclusion}
\begin{sloppypar}
By examining the equilibrated, zero-point fluctuations of the quantum Riemann tensor about some background, we can determine the fluctuation's probability distributions, entropy, and expected background.  This background obeys expectation-valued Einstein equations and features an entropy-based positive cosmological constant.  From the fluctuation probability distributions, a path integral can be constructed whose R-squared action describes quantum gravity in terms of `ordinary' quantum fields propagating on the `rigid' or `fixed' expected backgrounds.              
\end{sloppypar}
\section{Acknowledgments}
\begin{sloppypar}
I am grateful to Dr Stephen Fulling, Dr Chris Pope, and Mehmet Ozkan for their discussions and insightful questions.
\end{sloppypar}

\end{document}